\documentclass[aps,prl,twocolumn,showpacs,preprintnumbers,amsmath,amssymb,superscriptaddress]{revtex4}

\usepackage{graphicx}
\usepackage{dcolumn}

\newcommand{\be}{\begin{equation}}
\newcommand{\ee}{\end{equation}}
\newcommand{\bea}{\begin{eqnarray}}
\newcommand{\eea}{\end{eqnarray}}
\newcommand{\bm}[1]{\mathbf{#1}}

\newcommand{\lp}{\left(}
\newcommand{\rp}{\right)}

\def \cF{{\cal F}}

\def \cE{{\cal E}}
\def \cC{{\cal C}}
\def \cQ{{\cal Q}}

\newcommand{\ty}[1]{\mbox{\tiny #1}}

\begin{document}

\title{Electronic Cooling in Graphene}

\author{R. Bistritzer and A.H. MacDonald}
\affiliation{Department of Physics, The University of Texas at Austin, Austin Texas 78712\\}

\date{\today}

\begin{abstract}

Energy transfer to acoustic phonons is the dominant low-temperature cooling channel of electrons in a crystal.
For cold neutral graphene we find that the weak cooling power of its acoustical modes relative to the heat capacity of the system
leads to a power law decay of the electronic temperature when far from equilibrium.
For heavily doped graphene a high electronic temperature is shown to initially decrease linearly with time at a rate proportional to $n^{3/2}$ with $n$ being the electronic density.
We discuss the relative importance of optical and acoustic phonons to cooling.

\end{abstract}


\maketitle

\noindent
{\em Introduction}---Energy exchange between the electrons in nanoscale electronic devices and their environment
is a key issue in the design of electronic circuits and will play a role in any
future graphene-based electronics \cite{reviewMacDonald,reviewNeto}.  The dominant
electronic cooling mechanism in nearly any solid state environment is
energy transfer to phonons. Energy relaxation in a graphene sheet is dominated by transfer to the acoustic and optical
phonon modes of its two-dimensional honeycomb lattice and to the optical phonon modes of
its substrate \cite{Fuhrer,Guinea}.
In this Letter we address electronic energy relaxation in graphene with a focus on the intrinsic cooling channel provided
by the acoustical phonons.

Optical measurements are a particularly useful probe of energy transfer between electrons and
phonons and have been employed in the past in studies of electronic cooling in quantum wells\cite{ShahBook}.
Similar measurements were recently performed on epitaxial graphene samples\cite{deHeer,Spencer}.
In a typical measurement electrons are excited to high energies using an optical pulse.
The relaxation process of the hot electrons is then monitored using differential transmission spectroscopy.
Although transport measurements do in principle provide an alternate way of studying interactions between electrons and phonons\cite{Fuhrer},
the resistivity contribution from acoustic phonons in typical graphene samples  is much smaller
than the elastic-scattering residual resistivity contribution\cite{Morozov}.
Transport is therefore relatively insensitive to the electron-phonon coupling strength.
Even for suspended graphene samples in which
transport is nearly ballistic quantum resistance dominates over the phonon induced resistance\cite{Bolotin}.
On the contrary, cooling of hot Dirac quasi-particles is entirely due to phonons.

Guided by experiment\cite{deHeer,Spencer} we assume that e-e interactions thermalize the system throughout the relaxation process.
Given the lattice temperature $T_{\ty{L}}$ and the electronic density $n$, the cooling process is then characterized by a single time dependent function, the electronic temperature $T_e(t)$.
The time dependent chemical potential $\mu$ depends on $T_e$  and $n$ and is determined by the conservation of the number of particles.
As the hot electrons equilibrate the electronic temperature decreases approaching its
equilibrium value $T_{\ty{L}}$.

A unique situation arises for graphene in the {\em neutral regime}, $T_e \gg \mu$.
In a typical semiconductor this non-degenerate regime is reached only at high temperatures.
In nearly neutral graphene, on the other hand, this regime is accessed at nearly all temperatures of interest.
We find that when  $T_{\ty L} \ll T_e \lesssim 180K$ the electronic temperature satisfies a power-law decay law,
\be
T_e(t) = \frac{T_0}{\sqrt{t/\tau_0 + 1}},        \label{T_lowD_far_from_equilibrium}
\ee
with a characteristic time
\be
\tau_0 = \frac{424}{D^2 T_0^2} \ \mu{\rm s}.            \label{tau0}
\ee
Here $T_0$ in the initial temperature of the electrons and $D$ is the deformation potential measured in eV.
Hereafter we use $\hbar=1$ and measure all temperatures in meV.
Transport measurements have been able to bound the value of D between 10eV and 50eV however, more precise limits on this
important parameter have remained elusive\cite{Bolotin}.
We propose that Eq.(~\ref{T_lowD_far_from_equilibrium}) be used to obtain an accurate value of $D$ from experiment.

This slow low-$T_e$ cooling of neutral-regime Dirac quasi-particles differs markedly from the
very fast cooling observed when $T_e$ is high. Recent experimental\cite{deHeer,Spencer,Kampfrath} and theoretical\cite{Butscher} work
has demonstrated that very hot electron plasmas ($T_{e} \sim$ several hundred meV)
cool significantly after several pico-seconds.  We find that
low temperature cooling is slower by more than three orders of magnitude.

\noindent
{\em Theory of Temperature Dynamics}---
The assumption of rapid thermalization implies that
\be
\partial_t T_e = \cQ/\cC       \label{dt_Te_chain_rule}
\ee
where $\cC = \partial_{T_e} \cE$ is the electronic heat capacity, $\cQ = \partial_t \cE$ is the electronic cooling power \cite{james},
and $\cE$ is the energy density of the system.
Using the Boltzmann equation we find that
\be
\cQ = \partial_t \sum_{\bm{k}\alpha} \epsilon_{k\alpha} f_{\bm{k}}^{\alpha} = \sum_{\bm{k}\alpha} \epsilon_{k\alpha} S_{ph}(f_{\bm{k}}^{\alpha})      \label{Q_definition}
\ee
where
\be
S_{ph}(f_{\bm{k}\alpha}) = -\sum_{\bm{p}\beta}\left[ f_{\bm{k}}^{\alpha}\lp 1 - f_{\bm{p}}^{\beta} \rp W_{\bm{k}\alpha \to \bm{p}\beta} - \{\bm{k}\alpha \to \bm{p}\beta\} \right]      \label{S_ph}
\ee
is the collision integral.
Here $\alpha = v,c$ labels the valence and conduction bands
whose energies are $\pm v_g k$ with $v_g$ being graphene's band velocity.
The occupation of each band is given by the time dependent Fermi distribution function $f_{\bm{k}}^\alpha = f(\epsilon_{\bm{k}\alpha},T_e(t),\mu(t))$ and
\bea
W_{\bm{k} \alpha \to \bm{p} \beta} &=& 2\pi \sum_\bm{q} w^{\alpha\beta}_q \left[ (N_q+1) \delta_{\bm{k,p+q}} \delta(\epsilon^{\alpha\beta}_{\bm{kp}}-\omega_q)   \right. \nonumber \\
&+& \left. N_q \delta_{\bm{k,p-q}} \delta(\epsilon^{\alpha\beta}_{\bm{kp}}+\omega_q) \right]      \label{W}
\eea
is the transition rate between state $\bm{k}\alpha$ and state $\bm{p}\beta$. The energy exchanged with the phonon heat bath in the transition is
$\epsilon^{\alpha\beta}_{\bm{kp}} = \epsilon_{\bm{k}\alpha}-\epsilon_{\bm{p}\beta}$.
In Eq.(\ref{W}) $N_q = N(\omega_q)$ is the Bose distribution function evaluated at the phonon energy $\omega_q$.
For acoustical phonons the transition matrix element is $ w^{\alpha\beta}_q = D^2 q^2 (1 + s_{\alpha\beta} \cos\theta ) /4\rho \omega_q $.
Here $s_{\alpha\beta}=1$ for intraband transitions and $s_{\alpha\beta}=-1$ for the interband ones,
$\theta=\theta_k-\theta_p$ is the relative angle between the incoming and outgoing momenta and $\rho$ is the mass
density of graphene. For optical phonons $w^{\alpha\beta}_q \approx g^2$ where $g \approx 2v_g/a^2\sqrt{2\rho A \omega_0}$
with $a=1.42\AA$ and $A$ being the area of the graphene sheet\cite{Ando}.

We first consider $\cQ_a$ the energy transfer to the acoustic phonon bath which is accurately described by
the linear energy dispersion $\omega_q = c q$.
It is instructive to separate the energy transfer into a
loss due to spontaneous emission, $\cQ^{spont}$, and a gain due to
induced transitions, $\cQ^{ind}$.
Straightforward manipulations of (\ref{Q_definition}) lead to
\bea
\cQ^{ind} &=& -\frac{\pi D^2}{2\rho c} \sum_{\bm{k}\alpha \bm{p}\beta}  \epsilon^{\alpha\beta}_{\bm{kp}} \lp 1 + s_{\alpha\beta} \cos\theta \rp     \nonumber  \\
&\times& \lp f_{k}^{\alpha} - f_{p}^{\beta} \rp \sum_\bm{q} q N_q  \delta_{\bm{k,p+q}} \delta(\epsilon^{\alpha\beta}_{\bm{kp}}-\omega_q).    \label{Q_ind_1}
\eea
To evaluate $\cQ_a$ we exploit the large mismatch between $v_g$ and the sound velocity $c$ and evaluate
$\cQ_a$ to leading order in $c/v_g \ll 1$.  In the limit $c/v_g \to 0$ the scattering is elastic, only intraband transitions are allowed, and there is no energy loss.
For small $c/v_g$ interband scattering remains negligible and we can approximate $|\bm{p}|$ by $|\bm{k}|$
when performing the sum over $\bm{q}$ in Eq.(~\ref{Q_ind_1}).  In this way
we find that to $O(c/v_g)^4$
\be
\cQ^{ind} = \frac{D^2 T_L}{\rho v_g^2} \int \frac{k^3dk}{2\pi} \left[ f_k^c + (1 - f_k^v) \right].         \label{Q_ind_result}
\ee
Similar steps yield $Q^{spont}$ and the total energy loss
\be
\cQ_a = -\frac{D^2}{\rho v_g^2}\lp T_e - T_L \rp \int \frac{k^3 dk}{2\pi} \left[ f_k^c + (1 - f_k^v) \right].       \label{Q_result}
\ee
As expected the net energy loss vanishes when the electronic temperature reaches the lattice temperature.
Interestingly to leading order in $c/v_g$ the energy loss is independent of the sound velocity. Due to the absence
of inter-band transitions the cooling power from electrons in the conduction band and from holes in the valence band are simply additive.

As is evident from Eqs.(\ref{Q_ind_result}) and (\ref{Q_result}) the energy gain due to the induced transitions is negligible
compared to $\cQ^{spont}$ at high temperatures when $T_e \gg T_{\ty{L}}$. However as the system cools the difference between $\cQ^{ind}$
and $|\cQ^{spont}|$ decreases, vanishing in equilibrium. The vanishing of $\cQ$ in equilibrium is assured by the detailed balance condition
reflected by the collision integral expression, Eq.(~\ref{S_ph}).

The cooling power due to the intrinsic optical modes is easily estimated when the electron-phonon coupling is
approximated by a constant $g$ and phonon dispersion is neglected.
Setting the phonon energy to $\omega_0$ it follows from  Eqs.(\ref{Q_definition}--\ref{W}) that
\be
\cQ_o =  \frac{g^2 \omega_0^4}{(2\pi v_g^2)^2} [ N_e(\omega_0) - N_L(\omega_0) ] \cF(T_e,\mu)       \label{Qo}
\ee
where
\be
\cF(T_e,\mu) = \int_{-\infty}^{\infty} dx \; |x(x-1)| \; \left[f([x-1]\omega_0) - f(x\omega_0) \right].
\ee
Here $N_e$ and $N_{\ty L}$ are the Bose distribution functions evaluated
at the temperatures $T_e$ and $T_{\ty L}$ respectively, and the factor $|x(x-1)|$ originates from the
electronic joint density of states.  For neutral graphene $\cF \approx 1/6$ when $T_e \ll \omega_0$.
The dominant optical phonon bath in suspended graphene is likely intrinsic.
In unsuspended samples scattering by substrate phonons may introduce additional optical modes which are considerably less energetic\cite{Fuhrer}.
We focus below on the energy loss due to acoustical phonons and comment on the role of the intrinsic optical phonon modes only at the end of this manuscript.

\noindent
{\em Neutral limit}---
First we consider the neutral regime for which $\mu(t) \ll T_e(t)$.
If the equilibrium value of the chemical potential is significantly smaller than the lattice temperature
the system remains in the neutral regime throughout the entire relaxation process, otherwise the system will eventually exit the neutral regime as it approaches equilibrium.

The energy exchanged in a typical transition is $T_e$
implying a momentum transfer of $T_e/v_g$. The typical phonon energy is then $T_e c/v_g$ justifying the quasi-elastic approximation for
the scattering by acoustical phonons for \emph{all} values of the electronic temperature.
This situation is in marked contrast with the typical scenario in metals in which
e-ph scattering becomes highly inelastic
below the Bloch--Gr$\ddot{u}$neisen temperature.

In the neutral limit Eq.(~\ref{Q_result}) for $\cQ_a$ can be further simplified by
setting $\mu$ to zero in the  integral to obtain a value proportional to $T_{e}^4$.
Combining this result with Eq.(\ref{dt_Te_chain_rule}) and noting that
the energy density per spin and valley in neutral graphene is
$\cE = 3  \zeta(3) T_e^3/2\pi v_g^2$ where $\zeta$ is the Riemann zeta function we find that
\be
\partial_t T_e = -\gamma \; T_e^2 \; (T_e-T_{\ty{L}})      \label{dt_Te_lowD}
\ee
where $\gamma = 7\pi^4 D^2/540\zeta(3)\rho v_g^4 = 1.18\cdot 10^3 D^2 \ (meV^2\cdot sec)^{-1}$ with the deformation potential measured in eV's.
Both the cooling power and the heat capacity decrease as $T_{e}$ approaches $T_{\ty L}$. The cooling rate
slows because the former decrease is faster.
Far from equilibrium when $T_e \gg T_{\ty{L}}$ Eq.(\ref{dt_Te_lowD}) is solved by (\ref{T_lowD_far_from_equilibrium}).
The temperature decays as a power-law with a characteristic time of $\tau_0 = 1/2\gamma T_0^2$.
The simple relationship between $T_e$ and the energy density $\cE$ combined with Eq.(~\ref{T_lowD_far_from_equilibrium}) yield
\be
\cE(t) \approx   \frac{\cE_0}{\lp t/\tau_0 + 1 \rp^{3/2}}    \ \ \ \ \  \ \ T_e-T_{\ty{L}} \gg T_{\ty{L}}    \label{E_lowD}
\ee
where $\cE_0$ is the initial energy density of the system.
Like the electronic temperature, the energy density decays to equilibrium as a power-law when the system is far from equilibrium.

Near equilibrium, when $T_e \gtrsim {T_{\ty{L}}}$, we linearize Eq.(\ref{dt_Te_lowD}) with respect to $T_e-T_{\ty{L}}$ to find that
the electronic temperature decays exponentially to its equilibrium value with a characteristic time given by
$ \tau_L = 1/\gamma T_{\ty{L}}^2 = 848/D^2 T_{\ty{L}}^2 \mu {\rm s}$.


Our results for the neutral limit are valid to $O(\mu/T_e)^2$ since
the particle hole symmetry of the system implies that both
$\cQ_a$ and $\cE$ are even functions of $\mu/T_e$.
\begin{figure}[h]
\includegraphics[width=0.7\linewidth]{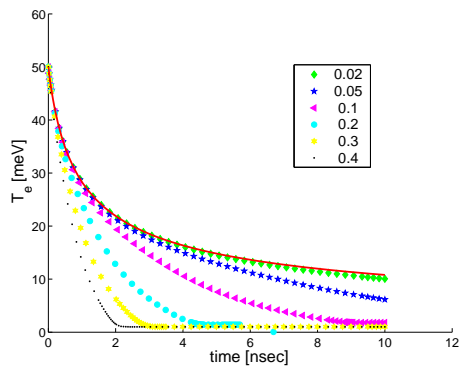}
\caption{(Color online) Equilibration of the electronic temperature. The evolution of the electronic temperature $T_e$ is plotted
for electronic densities of (up to down) 0.02, 0.05, 0.1, 0.2, 0.3 and 0.4 $\cdot 10^{12} [cm^{-2}]$. The lattice temperature is 1meV and the deformation potential
is assumed to be 20eV. The solid line corresponds to the equilibration of $T_e$ in a neutral system.}
\label{fig:Tn}
\end{figure}

\noindent
{\em Highly doped limit}---We now turn to study the equilibration of hot electrons in the doped regime for which $\mu(t) \gg T_e(t)$.
In this regime we use the Sommerfeld expansion to approximate $\cQ_a$ and $\cC$ and obtain
\be
\partial_t T_e = -\gamma_d \frac{T_e-T_{\ty L}}{T_e}
\ee
where $\gamma_d = 3 D^2 \epsilon_{\ty F}^3/4\pi^2 \rho v_g^4 = 0.133 D^2 n^{3/2}$meV/nsec with $n$ being measured in units of $10^{12} cm^{-2}$. Far from equilibrium
the high electronic temperature initially decreases linearly with time at an energy rate given by $\gamma_d$.
Near equilibrium $T_e$ approaches $T_{\ty L}$ exponentially at a rate of $\gamma_d/T_{\ty L}$.

The doped regime resembles the typical metallic case in that the quasi-elastic approximation
breaks down at low temperatures when the electronic temperature is below the  Bloch--Gr$\ddot{u}$neisen temperature $T_{\ty{}BG} = 2 c k_{\ty{F}}$. Therefore
our results for the doped regime describe the entire equilibration process for systems in which $T_{\ty{L}} > T_{\ty{BG}}$. However,
for cold doped graphene our results are valid only when $T_e(t) > T_{\ty{BG}}$.

\noindent
{\em General solution}---In the general case when $\mu(t) \sim T_e(t)$ the time evolution of the electronic temperature and chemical
potential follow from two coupled differential equations. The first equation
\be
0 = -2v_g \partial_t \lp 1/T_e \rp I_1^{(+)} + \partial_t \lp \mu/T_e \rp I_0^{(-)}       \label{number_eq_diff}
\ee
is obtained from the number equation $\partial_t n = 0$ and expresses the conservation of the number of particles throughout the relaxation process. Here
\be
I_n^{(\pm)} \equiv \int \frac{k^ndk}{2\pi} \left[ f_k^c \mp (1 - f_k^v) \right].          \label{I}
\ee
The second differential equation
\bea
&& -3v_g T_e \partial_t \lp 1/T_e \rp I_2^{(-)} + 2 T_e \partial_t \lp \mu/T_e \rp I_1^{(+)}       \nonumber \\
&=& -\frac{D^2}{\rho v_g^2} \lp T_e - T_{\ty{L}} \rp I_3^{(-)}.     \label{energy_eq_diff}
\eea
follows from Eqs.(\ref{dt_Te_chain_rule},\ref{Q_result}).

\begin{figure}[h]
\includegraphics[width=0.7\linewidth]{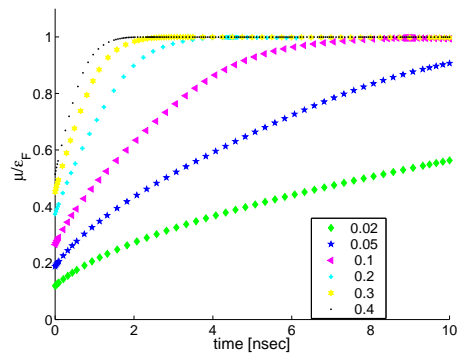}
\caption{(Color online) Equilibration of the chemical potential. The evolution of the chemical potential $\mu$ is scaled with the Fermi energy and plotted
for electronic densities of (from bottom to top) 0.02, 0.05, 0.1, 0.2, 0.3 and 0.4 $\cdot 10^{12} [cm^{-2}]$. The lattice temperature is 1meV and the deformation potential
is assumed to be 20eV.}
\label{fig:Mu_n}
\end{figure}

We have solved the coupled differential equations (\ref{number_eq_diff},\ref{energy_eq_diff}) numerically for various values of densities.
The electronic temperature as a function of time is plotted in Fig.\ref{fig:Tn} for $T_{\ty{L}} = 1meV$, $T_0 = 50 meV$ and $D=20 eV$.
Clearly the equilibration process is faster for doped systems.

The corresponding evolution of the chemical potential for the different electronic densities is plotted in Fig.\ref{fig:Mu_n}.
The conservation of particle number throughout the equilibration process forces $|\mu|$ to increase with time.

\begin{figure}[h]
\includegraphics[width=0.6\linewidth]{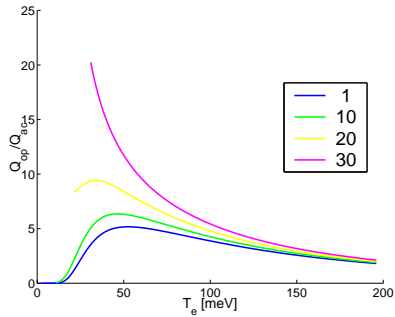}
\caption{The energy loss due to optical phonons divided by the energy loss do to acoustical phonons for neutral suspended
graphene is plotted vs. the electronic temperature for lattice
temperatures of (bottom to top) 1, 10, 20, and 30 meV.}
\label{fig:Qrel_neutral}
\end{figure}
\noindent
{\em Discussion}---
The small low-temperature cooling power of the neutral graphene electronic system is due to both
the small joint density-of-states for electronic transitions and the
small energy of acoustic phonons at typical transition momenta.
From dimension analysis of $\cQ$ and $\cE$ and
Eq.(\ref{dt_Te_chain_rule}) we find that the instantaneous energy decay rate $ \alpha = |\partial_t \ln(\cE)| = \cQ/|\cE|$
of a d-dimensional gapless
neutral system at zero lattice temperature with conduction and valence band dispersions $\epsilon_k \propto k^s$ satisfies
\be
\alpha \; \sim  \; T_e^{\frac{d-s+1}{s}}          \label{alpha_d_s}.
\ee
Exponential decay of $T_e$ and $\cE$ occurs when
$\alpha$ is constant, {\em i.e.} when $s=d+1$.
In systems like single layer or bilayer graphene with $s<d+1$, the cooling rate
will decrease as a power law of the instantaneous temperature when the system is far from equilibrium.
\begin{figure}[h]
\includegraphics[width=0.6\linewidth]{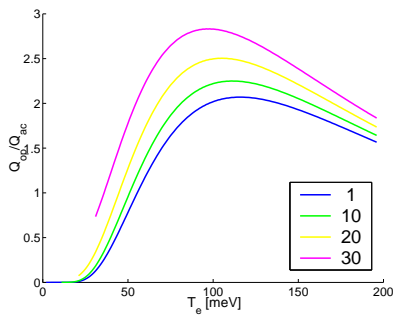}
\caption{The analogue of figure \ref{fig:Qrel_neutral} for a doped system with $n=10^{13} cm^{-2}$.}
\label{fig:Qrel_doped}
\end{figure}

It is interesting to consider the consequences of this work on the cooling of bilayer graphene.
The evolution of the electronic temperature of a gapless 2D system with a parabolic dispersion is given by $\partial_t T_e = -\tilde{\gamma} \sqrt{T_e} \lp T_e - T_{\ty{L}} \rp$.
However we do not expect this relationship to apply precisely in bilayer graphene since the momentum dependence of its energy spectrum can not be described by a single power \cite{Koshino}.

Although the main focus on this work has been energy loss due to acoustic phonons we emphasize that
the energetic optical phonons will play a dominant role in the cooling of graphene at sufficiently high
temperatures. To estimate the regime for which acoustic phonons dominate cooling in graphene we plot $\cQ_o/\cQ_a$ as a function
of $T_e$. We use a simple model in which $\omega_0=196 meV$ for both longitidinal and transverse optical branches.
In Fig.\ref{fig:Qrel_neutral} we plot $\cQ_o/\cQ_a$ for a neutral graphene sheet for various values
of $T_{\ty L}$. Surprisingly the different functional dependencies of $\cQ_o$ and $\cQ_a$ on the lattice and the
electronic temperatures lead to a non monotonic dependence of $\cQ_o/\cQ_a$ on $T_e$ and to the dominance
of $\cQ_o$ near equilibrium at moderate lattice temperatures.
As evident from Fig.\ref{fig:Qrel_doped} the relative cooling power of the acoustic phonons increases with doping,
increasing the maximum temperature at which they are dominant.

The inefficient cooling of graphene by acoustic phonons has immediate consequences for its non-linear transport\cite{Meric,us}.

\noindent
{\em Acknowledgment ---} This work has been supported by the Welch Foundation, by the Army Research Office,
by the NRI SWAN Center, and by the National Science Foundation under grant DMR-0606489. We acknowledge useful
discussions with Michael Fuhrer.

\end{document}